\documentclass[twocolumn,aps,prl,showpacs,amsmath,amssymb]{revtex4}

\usepackage{bm}
\newtheorem{theorem}{Theorem}

\begin{document}

\title[Stopping power of multi-component plasmas]{Fast projectile stopping power of quantal multi-component strongly coupled plasmas}
\author{D. Ballester}
\affiliation{School of Mathematics and Physics, Queen's University, Belfast BT7 1NN, United Kingdom}
\author{I. M. Tkachenko}
\affiliation{Departamento de Matem\'{a}tica Aplicada, Universidad Polit\'{e}cnica de Valencia, 46022 Valencia, Spain}%

\date{\today}

\begin{abstract}
The Bethe-Larkin formula for the fast projectile stopping power is extended
to multi-component plasmas. The results are to contribute to the correct
interpretation of the experimental data, which could permit to test the
existing and future models of thermodynamic, static, and dynamic
characteristics of strongly coupled Coulomb systems.
\end{abstract}

\maketitle


Stopping power is a characteristic of primary interest for different areas of physics such as
nuclear physics, condensed matter physics and plasma physics, as it arises when studying the
interaction of charged particles with matter. In 1930 Bethe derived his seminal formula for the
fast projectile energy losses assuming that the atoms of the medium behave as quantum-mechanical
oscillators \cite{bethe}. Later, Larkin \cite{L} showed that when fast ions permeate an electron
gas, an analogous formula is applicable, but with the mean excitation
frequency replaced by the plasma frequency $\omega _{p}$:%
\begin{equation}
-\frac{dE}{dx}\underset{v\gg v_{F}}{\simeq }\frac{4\pi Z_{p}^{2}e^{4}\rho N}{%
mAv^{2}} \ln \Lambda,  \label{L0}
\end{equation}%
where $\ln \Lambda= \ln 2mv^{2} /\hbar \omega _{p}$ is the quantal Coulomb logarithm, $Z_{p}e$ and
$v$ stand for the charge and velocity of the projectile, $\rho $ is the target density, $A$ the
mass of the target atoms, $N$ the Avogadro number, $v_{F}$ the electron Fermi velocity, and
$\omega _{p}=\left( 4\pi ne^{2}/m\right) ^{1/2}$, $m$ and $n$ being the electron mass and density. This formula is
usually employed to determine experimentally $n$ in a charged particle system. Particularly, its
applicability seems to be more promising in the field of plasma physics \cite{X1,X2,X3} for two
reasons: first, in an ionized medium the energy loss is mainly caused by the free electrons,
leading to an enhancement of the stopping power compared to the cold target \cite{X1,X2,X3};
secondly, this technique appears as the only suitable candidate for the diagnosis of hot and dense
($n\gtrsim 10^{19}$ $cm^{-3}$) plasmas, because most of the other methods fail under these conditions
\cite{X3}.

Usually, it is believed that the electronic subsystem of a plasma provides the main contribution
to the stopping power process, especially for fast projectiles. Our first aim in this Letter is to
show that in a multi-component completely ionized hydrogen plasma with a weakly damped Langmuir
mode of dispersion $\omega _{L}\left( k\right) $, the plasma frequency in the Coulomb logarithm of
(\ref{L0}) should be substituted by the long-wavelength limiting value of $\omega _{L}\left(
k\right) ,$ $\omega _{L}\left( 0\right) =\omega _{p}\sqrt{1+H}$ with $H=h_{ei}\left( 0\right)
/3=\left( g_{ei}\left( 0\right) -1\right) /3$, $g_{ei}\left( r\right) $ being the electron-ion
radial distribution function. The generalization to partially ionized plasmas or plasmas with
complex ions and more species is straightforward. This correction may have further practical
implications, in particular, after the experiments reported in Ref. \cite{X2} where it was
possible to measure separately the enhancement of the stopping power of fast ions due to the
increase in the Coulomb logarithm, $\ln\Lambda$. Thus, this method will permit to probe directly
strong coupling effects which are relevant to plasmas within the high density energy regime. This
includes plasmas arising in astrophysics and space science, planetary interiors, inertial
confinement fusion, matter under extreme conditions, metals and condensed matter plasmas.

Leaving the ionization losses aside, for calculating the stopping power for a fast projectile
passing through a Coulomb fluid we will adopt the polarizational picture, which becomes more
accurate as the kinetic energy of the projectile increases. In 1954 Lindhard obtained an
expression relating the polarizational stopping power with the medium (longitudinal) dielectric
function \cite{Li}. This expression can be generalized further by applying the Fermi golden rule
to obtain \cite{AB,BD93,OT01}:
\begin{equation}
-    \frac{dE}{dx}  =\frac{2\left(  Z_{p} e\right)  ^{2}}{\pi
v^{2}}\int\limits_{0}^{\infty}\frac{dk}{k}\int\limits_{\alpha_{-} \left( k \right ) } ^{\alpha_{+} \left( k \right )}
 \omega  n_{B}\left( \omega \right)  \left(- {\rm Im} \epsilon^{-1} \left(
k,\omega\right)  \right) d\omega  , \label{dedxpol1}%
\end{equation}
$\alpha_{\pm } \left( k \right ) = \pm kv +\hbar k^{2}/2M$, where $M$ is the mass of the
projectile (here we will work with heavy-ion projectiles, $M\gg m$), and $n_{B}=\left( 1-
\exp\left( -\beta \hbar \omega \right) \right)^{-1}$, $\beta^{-1}$ being the temperature in energy
units. In addition, unmagnetized Coulomb fluids are considered and, hence, the dielectric function
effectively depends only on the wavevector modulus. Expression (\ref{dedxpol1}) is valid only if
the interaction between the projectile and the plasma is so weak that it can be treated as a
linear effect and no relativistic effects need to be taken into account, i.e., when the energy
lost by a projectile is much less than its kinetic energy, which, in turn, is assumed to be much
smaller than its rest energy \footnote{In the experiments reported in Refs. \cite{X1,X2,X3} the
plasma temperature was of the order of a few eV (in Ref. \cite{X2} it is said to be below 500 eV),
whereas the projectiles were protons and deuterons at around 1 MeV.}.

The literature on the polarizational stopping power is very extensive. The problem has been
analyzed within the random-phase approximation (RPA) \cite{AB} and beyond, introducing an analytic
formula for the local field correction (LFC) factor \cite{44}. In addition there are also
nonlinear polarization effects \cite{barkas}, which are beyond the scope of this work. Whereas we
assume that the coupling between the projectile and the target plasma can be treated
perturbatively, we do not impose any restriction on the value of the coupling parameter,
$\Gamma=\beta e^{2} /a$ ($a= (4\pi n/3)^{-1/3}$ being the Wigner-Seitz radius), with the proviso
that the latter remains in the liquid phase \footnote{Strongly coupled plasmas are known to
crystallize at large values of coupling forming an anisotropic phase. See, e.g., M. Bonitz {\it et
al.}, Phys. Rev. Lett. 95, 235006 (2005), and references therein.}. As said before, here we will
focus on a completely ionized strongly coupled hydrogen plasma. The modeling of the dielectric
properties of this kind of plasmas constitutes a difficult problem, because its characteristic
lengths, i.e., Wigner-Seitz radius and Debye radius, $\lambda_{D}=(4 \pi n e^{2} \beta)^{-1/2}$,
are of the same order of magnitude (in a strongly coupled plasma $\Gamma = a^{2} / 3
\lambda_{D}^{2} \gtrsim 1$, what makes mean field theories, such as the RPA, and perturbative
treatments no longer valid) and, at the same time, its electronic subsystem is degenerate.

\paragraph{The framework.-\label{df}}

Our dielectric formalism is based on the method of moments \cite{amt,AT}, which allows to
determine the dielectric function $\epsilon(k,\omega)$ from the first known frequency moments or
sum rules. The sum rules we employ are actually the power frequency moments of the loss function
(LF) $\mathcal{L}\left(  k,\omega \right)  =- \omega ^{-1} \operatorname{Im}\epsilon ^{-1}\left(
k,\omega \right)  $
defined as
$C_{\nu}(k)= \pi ^{-1}\int_{-\infty}^{\infty} \omega^{\nu} \mathcal{L} \left(
k,\omega \right)  d\omega ,$ $\nu=0,1,\ldots $ . 
%
%
Due to the parity of the LF, all odd-order frequency moments vanish. The even-order frequency
moments are determined by the static characteristics of the system. After a straightforward
calculation one obtains \cite{kugler,amt,AT}:
$C_{0}(k)=(1-\epsilon^{-1}(k,0)) $, 
$C_{2}(k)=\omega_{p}^{2} $, and
$C_{4}(k)=\omega_{p}^{4} (1+K(k)+U(k)+H) $, 
with
$K(k)=  \left(  \left\langle v_{e}^{2}\right\rangle k^{2}+ \hbar^{2} k^{4} /
\left(  2m  \right)^{2}  \right) / \omega_{p}^{2} $, 
$\left\langle v_{e}^{2}\right\rangle $ being the average squared characteristic velocity of the
plasma electrons. The last two terms in $C_{4}$ can be expressed in terms of the partial structure
factors
$S_{ab}\left(  k\right)  ,$ $a,b=e,i$: %
$U\left(  k\right)  =\left( 2\pi^{2}n \right) ^{-1}%
{\int_{0}^{\infty}}
p^{2}\left(  S_{ee}\left(  p\right)  -1\right)  f\left(  p,k\right)  dp $,
$H=  \left( 6\pi^{2}n \right)^{-1}%
{\int_{0}^{\infty}}
p^{2}S_{ei}\left(  p\right)  dp $, 
where we have introduced %
$ f\left(  p,k\right)  = 5/12 - p^{2} / \left(4k^{2}\right) + \left(
k^{2}-p^{2}\right)  ^{2} \ln\left\vert \left(p+k \right) / \left( p-k \right)  \right\vert / \left( 8pk^{3} \right) $.

The Nevanlinna formula of the theory of moments expresses the dielectric function which satisfies
the known sum rules $\{ C_{2\nu} \}_{\nu=0}^{2}$ \cite{Nev,KN,amt}:
\begin{equation}
\epsilon^{-1}(k,z) = 1+\frac{\omega_{p}^{2}(z+q)}{z(z^{2}-\omega_{2}%
^{2})+q(z^{2}-\omega_{1}^{2})}, \label{24}%
\end{equation}
where $\omega_{1}^{2}=\omega_{1}^{2}\left(  k\right)  =C_{2}/C_{0}$,
$\omega_{2}^{2}=\omega_{2}^{2}\left(  k\right)  =C_{4}/C_{2} $, in terms of a function $q=q(k,z)$,
which is analytic in the upper complex half-plane $\operatorname{Im}\,z>0$ and has there a
positive imaginary part. It must also satisfy the limiting condition: $(q(k,z)/z)\rightarrow0$ as
$z\rightarrow\infty$ for $\operatorname{Im}\,z>0$. In an electron liquid this Nevanlinna parameter
function plays the role of the dynamic LFC $G\left( k,\omega\right)  $. In particular, the
Ichimaru visco-elastic model expression for $G\left( k,\omega\right)  $ is equivalent to the
Nevanlinna function approximated as $i/\tau_{m}$, $\tau_{m}$ being the effective relaxation time
of the Ichimaru model \cite{IchII}. In a multi-component
system the Nevanlinna parameter function stands for the species' dynamic LFC's. In general, we do
not have enough phenomenological conditions to determine that function $q(k,\omega)$ which would
lead to the exact expression for the LF. One might benefit from the Perel' - Eliashberg (PE)
\cite{PE} high-frequency asymptotic form \cite{amt}, $\mathrm{Im } \epsilon \left( k, \omega \gg
(\beta \hbar)^{-1} \right) \simeq \left( 4/3 \right)^{1/4} r_{s}^{3/4}/3 \left( \omega_{p} /
\omega \right)^{9/2} $,
where $r_{s}=a me^{2}/\hbar^{2}$ is the Brueckner parameter.

\paragraph{The corrected Bethe-Larkin formula.-}

Let us choose a model function $q$ satisfying the conditions mentioned after the Nevanlinna
formula (\ref{24}) that would permit to treat the stopping power calculation analytically. If we
put simply $q(k, \omega)=i0^{+}$, then we get the following particular solution of the moment
problem:
\begin{equation}
\frac{\mathcal{L} \left(  k,\omega\right)  }{\pi C_{0}\left(  k\right)  }%
=\frac{\omega_{2}^{2}-\omega_{1}^{2}}{\omega_{2}^{2}}\delta\left(
\omega\right)  +\frac{\omega_{1}^{2}}{2\omega_{2}^{2}}\left[  \delta\left(
\omega-\omega_{2}\right)  +\delta\left(  \omega+\omega_{2}\right)  \right]  ,
\label{can}%
\end{equation}
Physically, Eq. (\ref{can}) describes an undamped collective excitation mode (Feynman
approximation) at $ \omega_{2} $ with an additional central peak accounting for hydrodynamic
diffusional processes \cite{DK2003}. The applicability of this expression is justified provided
that the damping of the collective excitation is small enough, making this mode to act as the main
energy transfer channel. Thus we can disregard the details of the rest of the excitation spectrum.
If we introduce expression (\ref{can}) into the Lindhard formula (\ref{dedxpol1}),
it immediately reduces to:%
\begin{equation}
-\frac{dE}{dx} \underset{v\gg v_{F}}{\simeq}%
\frac{\left(  Z_{p}e\omega_{p}\right)  ^{2}}{v^{2}} \ln\frac{k_{2}}%
{k_{1}} , \label{dedxpol2}%
\end{equation}
where the "cut-off" wavenumbers $k_{1}$ and $k_{2}$ are such that the inequality $
0<\omega_{2}\left(  k \right)  < kv $ is satisfied with $v/v_{F} \rightarrow\infty$ and
$\omega_{2}\left(  k \right)  $ understood as the plasma Langmuir mode dispersion law
$\omega_{L}(k)$. For a weakly coupled plasma the RPA dispersion law is valid which neglects the
correlational contributions to $\omega_{L}(k)$:
$\omega_{L}(k)=\left( \omega_{p}^{2} + \left\langle
v_{e}^{2}\right\rangle k^{2} + \hbar^{2} k^{4} /
\left(  2m  \right)^{2}   \right)  ^{1/2}$. 
%
Then, if $v $ is asymptotically large, we have $ k_{1}=\omega_{p} /v$, $k_{2}=2mv/\hbar$, and we
recover the Bethe-Larkin (BL) result \cite{bethe,L}. Notice that in the above-mentioned inequality
for $\omega_{2}$, we have presumed that $kv\gg \hbar k^{2}/2M$, which is equivalent to disregard,
at most, terms of the order of $m/M$. Similar terms were omitted in the above expressions for the
moments $C_{2}$ and $C_{4}$, as well.

To take into account all Coulomb and exchange interactions in the system analytically, we might
use for the electron-electron contribution $U\left(k \right) $ its long- and short-range
asymptotic forms, $U\left( k\rightarrow0\right)  \simeq- v_{ee}^{2}k^{2} / \omega_{p}^{2}$,
$U\left(  k\rightarrow\infty\right)  \simeq- h_{ee}\left(  0\right) / 3 $, where $ v_{ee}^{2}=- 4
E_{ee}/ (15nm) $ is defined by the plasma electron-electron interaction energy density $E_{ee}$ of
the plasma \cite{OT01}, $h_{ee}\left(  0\right) $ being equal to the previous expression for
$U\left(k \right) $, but with the function $f(p,k)$ replaced by unity.
If we interpolate the plasma mode dispersion law as %
$\omega_{L}(k)= \left( \omega_{p}^{2}\left(1+ H\right) + w k^{2} %
+  \hbar^{2} k^{4} / \left(  2m  \right)^{2} \right)
^{1/2}$, %
with $ w =2\left\langle v_{e}^{2}\right\rangle -v_{ee}^{2}$, then
the "cut-off" wavenumber $k_{1}$ is modified as $
k^{\prime}_{1}=\omega_{p}^{\prime} / v $, with $\omega_{p}^{\prime}=
\omega_{p}\sqrt{ 1+ H}$, for $v/v_{F} \to \infty$, so
that the fast projectile stopping power becomes:%
\begin{equation}
- \frac{dE}{dx} \underset{v \gg v_{F} }{\simeq
} \left(  \frac{Z_{p}e\omega_{p}}{v}\right)  ^{2}\ln\frac{2m v^{2}  } {\hbar\omega_{p}\sqrt{1+ H }} .\label{L14}%
\end{equation}
Here, the correction $H$ stems from the electron-ion correlation contribution to the moment
$C_{4}(k)$ and is also the one responsible for the upshift in the value of the Langmuir frequency
predicted in the long-wavelength limit for an electron-ion plasma with an undamped collective
mode. Although the accurate calculation of $H$ under realistic conditions is a difficult task
\cite{HM,MP}, it is possible to find a simplified analytic expression based on the temperature
Green's function technique by a regularized summation over the Matsubara frequencies \cite{amt},
yielding $H=(4/3)r_{s}\sqrt{\Gamma}/(2\sqrt{r_s} +\Gamma\sqrt{6})$ (see also Ref. \cite{LeysMarch}
for an alternative approach based on a nontrivial renormalization via pair-correlations in liquid
metals). Whereas in a weakly coupled plasma, $\Gamma \ll 1$, this correction is negligible, in a
strongly coupled Coulomb system it could be possible to retrieve directly $H$ (or $g_{ei}(0)$) by
fitting Eq. (\ref{L14}) to some experimental data. For instance, if we take $g_{ei}(0)=10$
\cite{MP} and $\ln\Lambda=14$ \cite{X2}, then the stopping power obtained by the BL formula gets
modified by $\sim 5\%$, which indicates to what extent the experimental accuracy needs to be
improved.

\paragraph{The damped collective mode.-}

The collective mode is expected to be damped \cite{HM}, this implies that one cannot employ the
solution of the moment problem (\ref{can}) any longer. Here we will determine, on the basis of the
Chebyshev-Markov and other model-free inequalities, the bounds for the asymptotic form of the fast
projectile stopping power. Let us consider the contribution
\begin{eqnarray}
\mathcal{S}_{1} &:=&  \int_{0}^{k_{1}^{\prime\prime} \leq k_{1}^{\prime} } \frac{d k}{k}  \int_{\alpha_{-}}^{\alpha{+}}  \omega^{2}
n_{B} \left( \omega \right)  \mathcal{L} \left(  k, \omega \right) d\omega
 \nonumber \\
 & \leq &
  \int_{0}^{k_{1}^{\prime\prime} }
  \frac{4\pi^{2}e^{2}}{\hbar k^{3}} \alpha_{+} \left( k\right) dk \int_{0}^{\alpha_{+} }
   S\left( k,\omega\right) d{\omega} , \label{s1-1}
\end{eqnarray}
on account of the fluctuation-dissipation theorem (FDT) \cite{IchII}. Then, by applying the upper
bound obtained in Ref. \cite{MM} for the charge-charge static structure factor of a quantal multi-component
plasma under the assumption of perfect screening, $\lim_{k\to0}S(k)/k^{2}\leq  
\hbar\omega_p \coth ( \hbar\omega_{p} \beta/2) /(8\pi ne^{2})$,
we can approximate the previous integral as:
$\mathcal{S}_{1} \lesssim \pi \omega_{p} \coth ( \hbar\omega_{p} \beta/2 )
k_{1}^{\prime\prime} v  $, %
for $k_{1}^{\prime\prime} \leq k_{1}^{\prime} \sim v_{F} /v$.

This contribution should be compared with those stemming from
\begin{eqnarray}
\mathcal{S}_{2} &:=& \int_{k_{1}^{\prime\prime} \leq k_{1}^{\prime}}^{k_{2}^{\prime\prime} \geq k_{2} }
\frac{dk}{k} \int_{\alpha_{-}}^{\alpha_{+}} \omega^{2} n_{B} \left( \omega \right) \mathcal{L}
\left( k, \omega \right) d\omega \nonumber \\
 & \geq &   \int_{k_{1}^{\prime}}^{k_{2} }
\frac{dk}{k} \int_{0}^{kv} \omega^{2}  \mathcal{L} \left( k, \omega \right) d\omega .
\label{s2-1}
\end{eqnarray}
Clearly, we can find an upper bound for $\mathcal{S}_{2}$ analogous to expression (\ref{L14}). To
determine a lower bound we might apply the Chebyshev-Markov inequalities (CMI) \cite{KN}. In
particular, if we take the measure $d\sigma  = \omega^{2} \mathcal{L}  d\omega $, then
\begin{eqnarray}
\mathcal{S}_{2} \geq  \frac{\pi \omega_{p}^{2} }{2 } \int_{k_{1}^{\prime}}^{k_{2} } \frac{dk}{k}
\left( \frac{\left( kv \right)^{2} - \omega_{2}^{2}}{\left( kv \right)^{2} + \omega_{2}^{2}}
\right) . \label{s2-2}
\end{eqnarray}
Since we have assumed that $\forall k \in \left( k_{1}^{\prime}, k_{2}
\right)$, $kv
> \omega_{2}\left( k \right)$, then,
for some  $\xi>1$ such that $ \xi k_{1}^{\prime} \sim v_{F} / v $ as $v / v_{F}
\to \infty $,
we have%
\begin{eqnarray}
\mathcal{S}_{2} &\geq&  \frac{\pi \omega_{p}^{2} }{4 } \int_{k_{1}^{\prime}}^{\xi k_{1}^{\prime} }
\frac{dk}{k} \left( 1- \frac{\omega_{2}^{2}} {\left( kv \right)^{2}} \right) \nonumber \\
&=&  \frac{\pi \omega_{p}^{2} }{4 } \left( \ln \xi - \frac{\xi^{2}-1
}{2 \xi^{2}} \right)
+ \mathcal{O} \left( \frac{v_{F}^{2}}{v^{2}}
\right) . \label{s2-3}
\end{eqnarray}
Hence, if we want to ensure that the first leading term of the stopping power asymptote is
contained in $\mathcal{S}_{2}$, it is sufficient to choose the lower cut-off as
$k_{1}^{\prime\prime} \sim  v_{F}^{2} / v ^{2} $, to obtain $\mathcal{S}_{1} \leq \pi
\omega_{p}^{2} \coth ( \hbar\omega_{p} \beta/2 )  v_{F} / v  + \mathcal{O} \left( v_{F}^{2}
 /    v^{2}   \right) $, which becomes negligible compared to $S_{2}$ as $v / v_{F}
\to \infty $.

The last contribution to the stopping power (\ref{dedxpol1}) reads
\begin{equation}
\mathcal{S}_{3} := \int_{k_{2}^{\prime\prime} \geq k_{2}}^{\infty}
\frac{dk}{k}  \int_{\alpha_{-} }^{\alpha_{+}}  \omega^{2} n_{B}
\left( \omega \right) \mathcal{L} \left( k, \omega  \right) d\omega
.   \label{s3-1}
\end{equation}
In particular, if $k_{2}^{\prime\prime}\gg 2Mv/ \hbar = k_{2} M/m$, then
\begin{eqnarray}
\mathcal{I} \leq  \mathcal{S}_{3} \leq & n_{B} \left( \alpha_{-} \left( k_{2}^{\prime\prime}
\right) \right) \mathcal{I} , \label{s3-2}
\end{eqnarray}
$\mathcal{I}=\int_{k_{2}^{\prime\prime} }^{\infty} dk /k   \int_{\alpha_{-} }^{\alpha_{+}}
\omega^{2}
 \mathcal{L}  \left( k, \omega  \right)  d\omega$. By applying again the CMI, but now with the measure
$d\Sigma=\mathcal{L}d \omega$, it is possible to prove that the satisfaction of all three sum
rules, $\{C_{2\nu}\}_{\nu=0}^{2}$, alone does not guarantee the convergence of $\mathcal{S}_{3}$.
To this aim we may introduce an additional condition on the decay of the LF in the interval of
interest $\left( \alpha_{-},\alpha_{+}\right)$. Precisely, from the inequalities
\begin{equation}
\alpha_{-}^{2}   \int_{\alpha_{-} }^{\alpha_{+}}  d\Sigma \leq
\int_{\alpha_{-} }^{\alpha_{+}} \omega^{2} d\Sigma \leq
\alpha_{+}^{2} \int_{\alpha_{-} }^{\alpha_{+}}  d\Sigma ,
\label{s3-ineq}
\end{equation}
we see that $\mathcal{S}_{3}$ converges if and only if
$\int_{\alpha_{-}}^{\alpha_{+}} d \Sigma \lesssim \left( k_{F}/k
\right)^{\gamma}$, $\gamma>4$, which can be achieved by imposing on the
distribution $\Sigma \left( \omega \right)$ the following
H\"older condition:
\begin{equation}
\vert \Sigma \left( \alpha_{+} \right) - \Sigma \left( \alpha_{-}
\right) \vert \leq  \left( \frac{\omega_{p}}{\alpha_{+}}
\right)^{\mu} \left\vert \frac{\alpha_{+} - \alpha_{-} }{
\omega_{p}} \right\vert  ^{\nu} , \label{hoelder}
\end{equation}
with $0< \nu \leq 1$, $\mu \geq 3$, for $ k\geq
k_{2}^{\prime\prime}$. Then,
$ \mathcal{S}_{3} \lesssim 2 \omega_{p}^{2} M
k_{2} / \left(m k_{2}^{\prime\prime} \right) $. %
Therefore, it is feasible to choose an upper cut-off as $k_{2}^{\prime\prime} \sim  v^{2} /
v_{F}^{2} $ to get that $ \mathcal{S}_{3} \leq 2 \omega_{p}^{2}M  v_{F} / \left(m v\right) +
\mathcal{O} \left( v_{F}^{2} / v^{2} \right)$. A H\"older-type condition like (\ref{hoelder}) can
be fulfilled in a number of physical models, namely: in an electron-ion hydrogen plasma, where the
above-mentioned PE asymptote \cite{PE} is applicable if one assumes the spatial dispersion to be
negligible for wavelengths much higher than the maximum impact parameter, what is valid for the
range of frequencies and wavenumbers considered for $\mathcal{S}_{3}$. In case of a uniform
electron gas, the asymptotic expression derived in Ref. \cite{GL} satifies a similar condition as
well, although one needs to take into account the region of non-analyticity of the perturbative
expansion \cite{Holas}.

With the aforementioned conditions, it follows that:%
\begin{theorem}
The stopping power $-dE/dx$ given in (\ref{dedxpol1}) satisfies asymptotically, as $v/v_{F} \to
\infty$, $$\left( \frac{v_{F}}{v} \right)^{2}  \lesssim \left(  \frac{v_{F}} {Z_{p}e\omega_{p}}
\right)  ^{2}  \left(- \frac{dE}{dx} \right)  \lesssim \left( \frac{v_{F}}{v} \right)^{2} \ln
\frac{v}{v_{F}}. $$

\end{theorem}

This theorem provides the bounds for the fast projectile asymptotic form leading term. These are
based on inequalities which do not depend on the particular details of the fluctuation spectrum at
low and intermediate frequencies.

\paragraph{Conclusions.-}

In this Letter we have studied the modification of the BL expression for the plasma stopping power
due to the presence of an ion component, strong coupling and the decay of the Langmuir mode. We
have shown that, for a perfectly defined plasma collective mode with negligible damping, the
above-mentioned expression is affected by the electron-ion correlation. In addition, we have
derived bounds for the fast projectile asymptotic, on the basis of well-established results of the
linear response theory of Coulomb systems, namely, the zero-frequency sum rule, the f-sum rule,
the fourth moment sum rule, and the FDT, together with the compressibility sum rule. This general
result constitutes a sum rule for the calculation or numerical estimate of the fast projectile
stopping power for any model dielectric function satisfying the above-mentioned conditions, not
only in plasma physics, but also in other multi-component uniform charged particle models of condensed matter physics.

The authors acknowledge the financial support of the European Social Fund, the Spanish Ministerio
de Educaci\'on y Ciencia (Project ENE2007-67406-C02-02/FTN), and the INTAS (Project
06-1000012-8707).

\end{document}